\documentclass[conference, 10pt]{IEEEtran}
\usepackage[dvips]{graphicx}
\usepackage{amssymb,amsmath}
\usepackage{subfig}
\usepackage{colortbl}
\usepackage{url}

\makeatletter
\def\ps@headings{%
\def\@oddhead{\mbox{}\scriptsize\rightmark \hfil \thepage}%
\def\@evenhead{\scriptsize\thepage \hfil \leftmark\mbox{}}%
\def\@oddfoot{}%
\def\@evenfoot{}}
\makeatother
\IEEEoverridecommandlockouts

\newcommand{\ie}{\emph{i.e. }}
\newcommand{\etal}{\textit{et al.}}
\newcommand{\n}[1]{\mathcal{N}_{#1}}
\newcommand{\thru}[1]{\mathcal{T}_{#1}}

\begin{document}

\title{Modeling Network Coded TCP Throughput:\\ A Simple Model and its Validation}
\author{
\authorblockN{MinJi Kim\authorrefmark{1}, Muriel M\'{e}dard\authorrefmark{1}, Jo\~{a}o Barros\authorrefmark{2}\vspace*{.5cm}}
\hspace*{2cm} \authorrefmark{1}{\normalsize Research Laboratory of Electronics} \hspace*{3.3cm} \authorrefmark{2}{\normalsize Instituto de Telecommunica\c{c}\~{o}es}\hfill\\
\hspace*{2cm} {\normalsize Massachusetts Institute of Technology} \hspace*{1cm} {\normalsize Departamento de Engenharia Electrot\'{e}cnica e de Computadores}\hfill\\
\hspace*{2.5cm}{\normalsize Cambridge, MA 02139, USA} \hspace*{2cm} {\normalsize Faculdade de Engenharia da Universidade do Porto, Portugal}\hfill\\
\hspace*{2.2cm}{\normalsize Email: \{minjikim, medard\}@mit.edu} \hspace*{4cm} {\normalsize Email: jbarros@fe.up.pt}\hfill
}


\maketitle

\begin{abstract}
We analyze the performance of TCP and TCP with network coding (TCP/NC) in lossy wireless networks. We build
upon the simple framework introduced by Padhye \etal\ and characterize the throughput behavior of classical TCP as well as TCP/NC as a function of erasure rate, round-trip time, maximum window size, and duration of the connection. Our analytical results show that network coding masks erasures and losses from TCP, thus preventing TCP's performance degradation in lossy networks, such as wireless networks. It is further seen that TCP/NC has significant throughput gains over TCP. In addition, we simulate TCP and TCP/NC to verify our analysis of the average throughput and the window evolution. Our analysis and simulation results show very close concordance and support that TCP/NC is robust against erasures. TCP/NC is not only able to increase its window size faster but also to maintain a large window size despite losses within the network, whereas TCP experiences window closing essentially because losses are mistakenly attributed to congestion.
\end{abstract}
\IEEEpeerreviewmaketitle

\section{Introduction}\label{sec:introduction}

The Transmission Control Protocol (TCP) is one of the core protocols of today's Internet Protocol Suite. TCP was designed for reliable transmission over wired networks, in which losses are generally indication of congestion. This is not the case in wireless networks, where losses are often due to fading, interference, and other physical phenomena. In wireless networks, TCP often incorrectly assumes that there is congestion within the network and unnecessarily reduces its transmission rate, when it should have actually transmitted continuously to overcome the lossy links. Consequently, TCP's performance in wireless networks is poor when compared to the wired counterparts as shown e.g. in \cite{caceres}\cite{TCP_Kurose}. There has been extensive research to combat these harmful effects of erasures and failures; however, TCP even with modifications does not achieve significant improvement. References \cite{hari}\cite{Tian05tcpin} give an overview and a comparison of various TCP versions over wireless links.

Some relief may come from network coding \cite{ahlswede}, which has been introduced as a potential paradigm to operate communication networks, in particular wireless networks. Network coding allows and encourages mixing of data at intermediate nodes, which has been shown to increase throughput and robustness against failures and erasures \cite{algebraic}. There are several practical protocols that take advantage of network coding in wireless networks. For example, opportunistic coding schemes with linear network coding are proposed in \cite{xor}\cite{more}\cite{codeOr}\cite{costa}.

\begin{figure}[tbp]
\begin{center}\vspace*{.4cm}
\includegraphics[width=.44\textwidth]{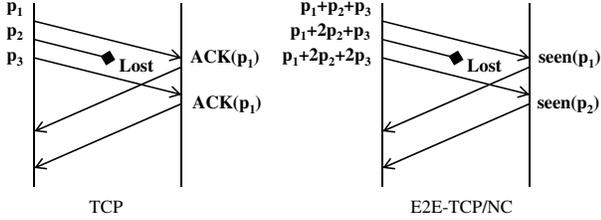}
\end{center}\vspace*{-.3cm}\caption{Example of TCP and E2E-TCP/NC. In the case of TCP, the TCP sender receives duplicate ACKs for packet $\mathbf{p_1}$, which may wrongly indicate congestion. However, for E2E-TCP/NC, the TCP sender receives ACKs for packets $\mathbf{p_1}$ and $\mathbf{p_2}$; thus, the TCP sender perceives a longer round-trip time (RTT) but does not mistake the loss to be congestion.}\label{fig:example}\vspace*{-.4cm}
\end{figure}

In order to combine the benefits of TCP and network coding, \cite{tcpnc} proposes a new protocol called TCP/NC. The key idea is a new network coding layer between the transport layer and the network layer, which incurs minimal changes to the protocol stack. TCP/NC modifies TCP's acknowledgement (ACK) scheme such that it acknowledges \emph{degrees of freedom} instead of individual packets, as shown in Figure \ref{fig:example}. This is done so by using the concept of ``seen'' packets -- in which the number of degrees of freedom received is translated to the number of consecutive packets received. In \cite{tcpnc}, the authors present two versions of TCP/NC -- one that adheres to the end-to-end philosophy of TCP, in which encoding and decoding operations are only performed at the source and destination, and another one that takes advantage of network coding even further by allowing any subset of intermediate nodes to re-encode. Note that re-encoding at the intermediate nodes is an optional feature of TCP/NC, and is not required for TCP/NC to work.

In this paper, we shall focus on TCP as well as TCP/NC with end-to-end network coding, which we denote E2E-TCP/NC (or in short E2E), in lossy networks. We adopt the same TCP model as in \cite{TCP_Kurose} -- \ie we consider standard TCP with Go-Back-N pipelining. Thus, the standard TCP discards packets that are out-of-order. We analytically show the throughput gains of E2E over standard TCP, and present simulations results that support this analysis. We develop upon the model introduced in \cite{TCP_Kurose} to characterize the steady state throughput behavior of both TCP and E2E as a function of erasure rate, round-trip time (RTT), and maximum window size. Our work thus extends the work of \cite{TCP_Kurose} for E2E-TCP/NC in lossy wireless networks. Furthermore, we use NS-2 (Network Simulator \cite{ns}) to verify our analytical results for TCP and E2E. Our analysis and simulations show that E2E is robust against erasures and failures. E2E is not only able to increase its window size faster but also maintain a large window size despite losses within the network. Thus, E2E is well suited for reliable communication in lossy networks. In contrast, standard TCP experiences window closing as losses are mistaken to be congestion.

The paper is organized as follows. In Section \ref{sec:model}, we introduce our communication model. In Section \ref{sec:td-intuition}, we briefly provide the intuition behind the benefit of using network coding with TCP. Then, we provide throughput analysis for TCP and E2E in Sections \ref{sec:td} and \ref{sec:tcpnc}, respectively. In Section \ref{sec:down}, we discuss the throughput behavior when the network is experiencing severe congestion, deep fading, and/or adversarial jamming. In Section \ref{sec:simulations}, we provide simulation results to verify our analytical results in Sections \ref{sec:td} and \ref{sec:tcpnc}.
Finally, we conclude in Section \ref{sec:conclusions}.

\section{A Model for TCP Congestion Control}\label{sec:model}
We focus on TCP's congestion avoidance mechanism, where the congestion control window size $W$ is incremented by $1/W$ each time an ACK is received. Thus, when every packet in the congestion control window is ACKed, the window size $W$ is increased to $W+1$. On the other hand, the window size $W$ is reduced whenever an erasure/congestion is detected.

We model TCP's behavior in terms of \emph{rounds} as in \cite{TCP_Kurose}. We denote $W_i$ to be the size of TCP's congestion control window size at the beginning of round $i$. The sender transmit $W_i$ packets in its congestion window at the start of round $i$, and once all $W_i$ packets have been sent, it defers transmitting any other packets until at least one ACK for the $W_i$ packets are received. The ACK reception ends the current round, and starts round $i+1$.

For simplicity, we assume that the duration of each round is equal to a round trip time ($RTT$), independent of $W_i$. This assumes that the time needed to transmit a packet is much smaller than the round trip time. This implies the following sequence of events for each round $i$: first, $W_i$ packets are transmitted. Some packets may be lost.
The receiver transmits ACKs for the received packets. (Note that TCP uses cumulative ACKs. Therefore, if the packets $1, 2, 3, 5, 6, 7$ arrive at the receiver in sequence, then the receiver ACKs packets $1, 2, 3, 3, 3, 3$. This signals that it has not yet received packet 4.) Some of the ACKs may also be lost. Once the sender receives the ACKs, it updates its window size. Assume that $a_i$ packets are acknowledged in round $i$. Then, $W_{i+1} \leftarrow W_i + a_i/W_i$.

TCP reduces the window size for congestion control using the following two methods.
\begin{enumerate}
\item \emph{Triple-duplicate (TD):} When the TCP sender receives four ACKs with the same sequence number, then $W_{i+1} \leftarrow \frac{1}{2} W_i$.
\item \emph{Time-out (TO):} If the Sender does not hear from the receiver for a predefined time period, called the ``time-out'' period (which is $T_o$ rounds long), then the sender closes its transmission window, $W_{i+1} \leftarrow 1$. At this point, the sender updates its TO period to $2T_o$ rounds, and transmits one packet. For any subsequent TO events, the sender transmits the one packet within its window, and doubles its TO period until $64T_o$ is reached, after which the time-out period is fixed to $64T_o$. Once the sender receives an ACK from the receiver, it resets its TO period to $T_o$ and increments its window according to the congestion avoidance mechanism. During time-out, the throughput of both TCP and E2E is zero.
\end{enumerate}

Finally, we note that in practice, the TCP receiver sends a single cumulative ACK after receiving $\beta$ number of packets, where $\beta = 2$ typically. However, we assume that $\beta = 1$ for simplicity. Extending the analysis to $\beta \geq 1$ is straightforward.

\subsection{Maximum window size}

In general, TCP cannot increase its window size unboundedly; there is a maximum window size $W_{\max}$. The TCP sender uses a congestion avoidance mechanism to increment the window size until $W_{\max}$, at which the window size remains $W_{\max}$ until a TD or a TO event.

\subsection{Erasures}

We assume that there are two different states: \emph{up-state} and \emph{down-state}. The network is in down-state when the network fails and the sender times-out. This may occur due to severe congestion, adversarial jamming, interference and deep fading, which are especially relevant for wireless networks. We denote $p_{d}$ to be the probability that the network is in the down-state during any given round. Note that during down-state, both TCP and E2E-TCP/NC have throughput of zero, as the forward and/or the backward paths have failed.

Assuming that the network is in up-state, we denote $p$ to be the probability that a packet is lost at any given time. We further assume that packet losses are independent.
We note that this erasure model is different from that of \cite{TCP_Kurose} where losses are correlated within a round -- \ie bursty erasures. Correlated erasures model well bursty traffic and congestion in wireline networks. In our case, however, we are aiming to model wireless networks, thus we shall use random independent erasures.


%
%


\subsection{Performance metric}

\begin{figure*}[tbp]
\begin{center}
\subfloat[TCP]{\label{fig:td-tcp}\includegraphics[width=.40\textwidth]{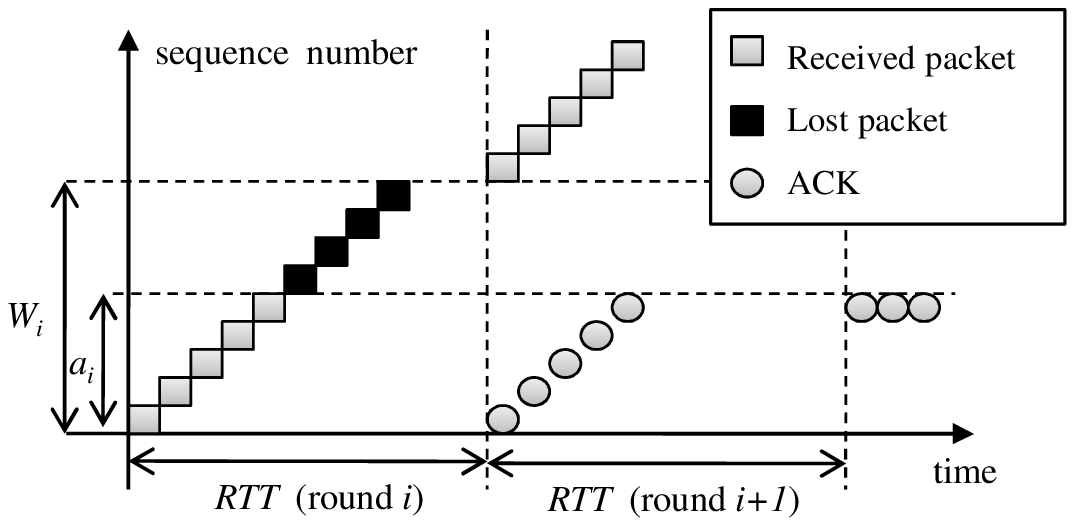}}
\hspace*{.7cm}
\subfloat[E2E-TCP/NC]{\label{fig:td-tcpnc}\includegraphics[width=.40\textwidth]{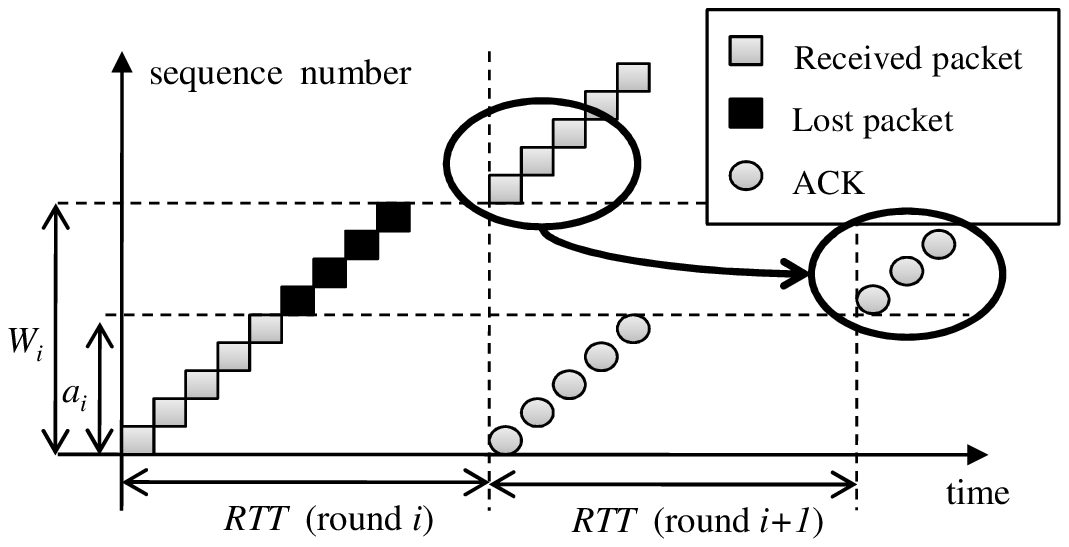}}
\end{center}\vspace*{-.3cm}
\caption{The effect of erasures: TCP experiences triple-duplicate ACKs, and results in $W_{i+2} \leftarrow W_{i+1}/2$. However, E2E-TCP/NC masks the erasures using network coding, which allows TCP to advance its window. This figure depicts the sender's perspective, therefore, it indicates the time at which the sender transmits the packet or receives the ACK.}\vspace*{-.3cm}\label{fig:td}
\end{figure*}\label{fig:td}

We analyze the performance of TCP and E2E in terms of two metrics: the average throughput $\thru{}$, and the expected window evolution $E[W]$, where $\thru{}$ represents the total average throughput while window evolution $E[W]$ reflects the perceived throughput at a given time.
We define $\n{[t_1, t_2]}$ to be the number of packets received by the receiver during the interval $[t_1,t_2]$. The total average throughput is defined as:
\begin{equation}\label{eq:thru-era}
\thru{} = \lim_{\Delta \rightarrow \infty} \frac{\n{[t, t+\Delta]}}{\Delta}.
\end{equation}
We denote $\thru{tcp}$ and $\thru{e2e}$ to be the average throughput for TCP and E2E, respectively.


\section{Intuition}\label{sec:td-intuition}

For traditional TCP, erasures in the network can lead to triple-duplicate ACKs. For example, in Figure \ref{fig:td-tcp}, the sender transmits $W_i$ packets in round $i$; however, only $a_i$ of them arrive at the receiver. As a result, the receiver ACKs the $a_i$ packets and waits for packet $a_i + 1$. When the sender receives the ACKs, round $i+1$ starts. The sender updates its window ($W_{i+1} \leftarrow W_i + a_i/W_i$), and starts transmitting the new packets in the window. However, since the receiver is still waiting for packet $a_i+1$, any other packets cause the receiver to request for packet $a_i +1$. This results in a triple-duplicate ACKs event and the TCP sender closes its window, \ie $W_{i+2} \leftarrow \frac{1}{2} W_{i+1} = \frac{1}{2} (W_{i} + a_i/W_i)$.

Notice that this window closing due to TD does not occur when using E2E as illustrated in Figure \ref{fig:td-tcpnc}. This is due to the fact that, with network coding, any linearly independent packet delivers new information. Thus, any subsequent packet (in Figure \ref{fig:td-tcpnc}, the first packet sent in round $i+1$) can be viewed as packet $a_i + 1$. As a result, the receiver is able to increment its ACK and the sender continues transmitting data.

It follows that network coding masks the losses within the network from TCP, and prevents it from closing its window by misjudging link losses as congestion. It is important to note that {\bf\emph{network coding translates losses as longer RTT}}, thus slowing down the transmission rate to adjust for losses without closing down the window in a drastic fashion.

\section{Throughput Analysis for TCP}\label{sec:td}

In this section, we consider the effect of losses for TCP. The throughput analysis for TCP is similar to that of \cite{TCP_Kurose}. However, the model has been modified from that of \cite{TCP_Kurose} to account for independent erasures and allow a fair comparison with network coded TCP. Assuming that the network is in its up-state, TCP can experience a TD or a TO event. As in \cite{TCP_Kurose}, we first consider TD events, and then incorporate TO events.

We note that, despite independent packet erasures, a single packet loss may affect subsequent packet reception. This is due to the fact that TCP requires in-order reception. Thus, a single packet loss within a transmission window forces all subsequent packets in the window to be out of order. Thus, they are discarded by the TCP receiver. As a result, standard TCP's throughput behavior with independent losses is similar to that of \cite{TCP_Kurose}, where losses are correlated within one round.


\begin{figure}[tbp]
\begin{center}\hspace*{-.3cm}
\includegraphics[width=0.46\textwidth]{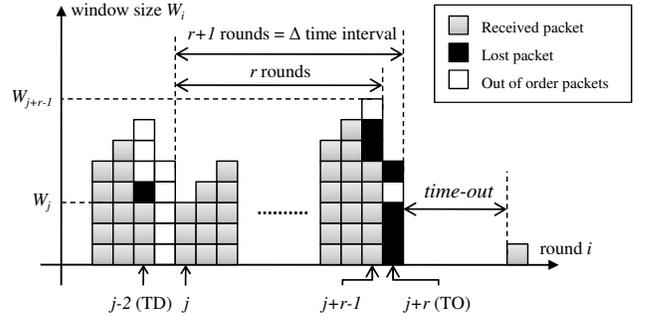}
\end{center}\vspace*{-.3cm}\caption{TCP's window size with a triple-duplicate ACKs event and a time-out event. In round $j-2$, losses occur resulting in triple-duplicate ACKs. On the other hand, in round $j+r-1$, losses occur; however, in the following round $j+r$ losses occur such that the TCP sender only receives two-duplicate ACKs. As a result, TCP experiences time-out.}\vspace*{-.3cm}\label{fig:td-rounds}
\end{figure}

\subsection{Triple-duplicate for TCP}\label{sec:td-tcp}

We consider the expected throughput between consecutive TD events, as shown in Figure \ref{fig:td-rounds}. Assume that the TD events occurred at time $t_1$ and $t_2 = t_1 + \Delta$, $\Delta > 0$. Assume that round $j$ begins immediately after time $t_1$, and that packet loss occurs in the $r$-th round, \ie round $j+r-1$.

First, we calculate $E[\n{[t_1, t_2]}]$. Note that during the interval $[t_1, t_2]$, there are no packet losses. Given that the probability of a packet loss is $p$, the expected number of consecutive packets that are successfully sent from sender to receiver is
\begin{equation}\label{eq:td-packet}
E\left[\n{[t_1, t_2]}\right] = \left(\sum_{k=1}^\infty k (1-p)^{k-1} p \right)- 1= \frac{1-p}{p}.
\end{equation}

The packets (in white in Figure \ref{fig:td-rounds}) sent after the lost packets (in black in Figure \ref{fig:td-rounds}) are out of order, and will not be accepted by the standard TCP receiver. Thus, Equation (\ref{eq:td-packet}) does not take into account the packets sent in round $j-1$ or $j+r$.

We calculate the expected time period between two TD events, $E[\Delta]$. As in Figure \ref{fig:td-rounds}, after the packet losses in round $j$, there is an additional round for the loss feedback from the receiver to reach the sender. Therefore, there are $r+1$ rounds within the time interval $[t_1, t_2]$, and $\Delta = RTT ( r + 1)$. Thus,
\begin{equation}\label{eq:td-time}
E[\Delta] = RTT (E[r]+1).
\end{equation}
To derive $E[r]$, note that $W_{j+r-1} = W_{j}+ r-1$ and
\begin{equation}\label{eq:halfwindow}
W_j = \frac{1}{2} W_{j-1} = \frac{1}{2} \left(W_{j-2} + \frac{a_{j-2}}{W_{j-2}}\right).
\end{equation}
Equation (\ref{eq:halfwindow}) is due to TCP's congestion control. TCP interprets the losses in round $j-2$ as congestion, and as a result halves its window. Assuming that, in the long run, $E[W_{j+r-1}] = E[W_{j-2}]$ and that $a_{j-2}$ is uniformly distributed between $[0, W_{j-2}]$,
\begin{equation}\label{eq:td-window1}
E[W_{j+r-1}] = 2 \left(E[r] -\frac{3}{4}\right) \text{  and  } E[W_j] = E[r] -\frac{1}{2}.
\end{equation}
During these $r$ rounds, we expect to successfully transmit $\frac{1-p}{p}$ packets as noted in Equation (\ref{eq:td-packet}). This results in the following equations:
\begin{align}
\frac{1-p}{p}& = \sum_{k=0}^{r-1} a_{j+k} = \left(\sum_{k=0}^{r-2} W_{j+k}\right) + a_{j+r-1}\label{eq:td-2}\\
& = (r-1)W_j + \frac{(r-1)(r-2)}{2} + a_{j+r-1}. \label{eq:td-1}
\end{align}
Taking the expectation of Equation (\ref{eq:td-1}) and using Equation (\ref{eq:td-window1}),
\begin{equation}\label{eq:quad}
\frac{1-p}{p} = \frac{3}{2}(E[r]-1)^2 + E[a_{j+r-1}].
\end{equation}
Note that $a_{j+r-1}$ is assumed to be uniformly distributed across $[0, W_{j+r-1}]$. Thus, $E[a_{j+r-1}] = E[W_{j+r-1}]/2 = E[r]-\frac{3}{4}$ by Equation (\ref{eq:td-window1}). Solving Equation (\ref{eq:quad}) for $E[r]$, we get the following:
\begin{equation}\label{eq:td-tcp-r}
E[r] = \frac{2}{3} + \sqrt{-\frac{1}{18} + \frac{2}{3}\frac{1-p}{p}}.
\end{equation}
This provides an expression of steady state average window size for TCP (using Equations (\ref{eq:td-window1}) and (\ref{eq:td-tcp-r})):
\begin{align}
E[W] &= \frac{E[W_j] + E[W_{j+r-1}]}{2}\\
&=\frac{3}{2}E[r] -1. \label{eq:tcp-w}
\end{align}
The average throughput can be expressed as
\begin{equation}\label{eq:td-tcp-thru}
\thru{tcp}' = \frac{E[\n{[t_1, t_2]}]}{E[\Delta]} = \frac{1-p}{p}\frac{1}{RTT(E[r]+1)}.
\end{equation}
For small $p$, $\thru{tcp}' \approx \frac{1}{RTT}\sqrt{\frac{3}{2p}} + o(\frac{1}{\sqrt{p}})$; for large $p$, $\thru{tcp}' \approx \frac{1}{RTT}\frac{1-p}{p}$. If we only consider TD events, the long-term steady state throughput is equal to that in Equation (\ref{eq:td-tcp-thru}).

The analysis above assumes that the window size can grow unboundedly; however, this is not the case. To take maximum window size $W_{\max}$ into account, we make a following approximation:
\begin{equation}\label{eq:td-tcp}
\thru{tcp} = \min\left(\frac{W_{\max}}{RTT}, \thru{tcp}'\right).
\end{equation}
For small $p$, this result coincide with the results in \cite{TCP_Kurose}.

\subsection{Time-out for TCP}

We note that even when the network is in the up-state, TCP can experience TO events. This happens when enough loss events occur within two consecutive rounds such that there are only two or fewer out of order packets and all others are lost, as shown in Figure \ref{fig:td-rounds}.
Thus, $\mathbf{P}(\text{TO}|W)$, the probability of a TO event given a window size of $W$, is given by
\begin{equation}\label{eq:probTO}
\small
\mathbf{P}(\text{TO}|W) =
\begin{cases}
1 &\text{if $W <3$;}\\
\sum_{i=0}^2 \binom{W}{i} p^{W-i}(1-p)^i &\text{if $W \geq 3$}.
\end{cases}
\end{equation}

\begin{figure*}[tbp]
\small
\begin{align}
E[\text{duration of TO period}] &= (1-p) \left[T_o p  + 3T_o p^2 + 7 T_o p^3  + 15 T_o p^4 + 31 T_o p^5  + \sum_{i=0}^{\infty} (63 + i\cdot 64)T_o p^{6 + i}\right]\\
&= (1-p) \left[T_o p  + 3T_o p^2 + 7 T_o p^3  + 15 T_o p^4 + 31 T_o p^5  + 63T_o\frac{p^6}{1-p} + 64T_o\frac{p^7}{(1-p)^2} \right] \label{eq:down}
\end{align}\vspace*{-.3cm}
\end{figure*}

\begin{figure*}[tbp]
\small
\begin{equation}\label{eq:tcp}
\thru{tcp} = \min\left(\frac{W_{\max}}{RTT}, \frac{1-p}{p}\frac{1}{RTT\left(\frac{5}{3} + \sqrt{-\frac{1}{18} + \frac{2}{3}\frac{1-p}{p}} + \mathbf{P}(\text{TO}|E[W])E[\text{duration of TO period}]\right)} \right)
\end{equation}\vspace*{-.3cm}
\end{figure*}

We approximate $W$ in above Equation (\ref{eq:probTO}) with the expected window size $E[W]$ from Equation (\ref{eq:tcp-w}). The length of the TO event depends on the duration of the loss events. Thus, the expected duration of TO period (in RTTs) is given in Equation (\ref{eq:down}). Finally, by combining the results in Equations (\ref{eq:td-tcp}), (\ref{eq:probTO}), and (\ref{eq:down}), we get an expression for the average throughput of TCP as shown in Equation (\ref{eq:tcp}).


\section{Throughput Analysis for E2E-TCP/NC}\label{sec:tcpnc}

\begin{figure}[tbp]
\begin{center}
\includegraphics[width=0.47\textwidth]{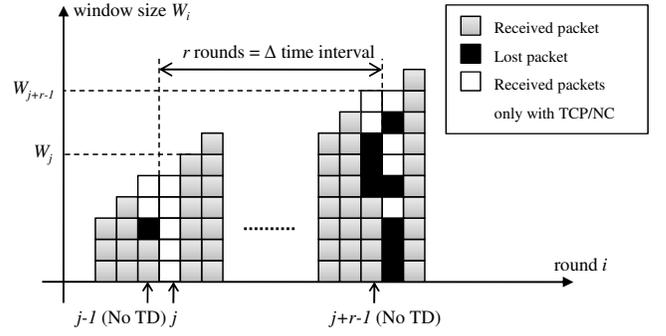}
\end{center}\vspace*{-.3cm}\caption{E2E-TCP/NC's window size with erasures that would lead to a triple-duplicate ACKs event when using standard TCP. Note that unlike TCP, the window size is non-decreasing.}\label{fig:td-tcpncrounds}\vspace*{-.3cm}
\end{figure}

We consider the expected throughput for E2E given that the network is in up-state. It is important to note that erasure patterns that result in TD and/or TO events under TCP may not yield the same result under E2E, as illustrated in Section \ref{sec:td-intuition}. We emphasize again that this is due to the fact that any linearly independent packet conveys a new degree of freedom to the receiver. Figure \ref{fig:td-tcpncrounds} illustrates this effect -- packets (in white) sent after the lost packets (in black) are acknowledged by the receivers, thus allowing E2E to advance its window.

This implies that E2E does not experience window closing often during the network up-state. We shall show that the window size $W_i$ is actually a non-decreasing function in $i$. However, not surprisingly, the rate at which $W_i$ grows depends on $p$ as we shall show in the subsequent sections.


\subsection{E2E-TCP/NC Window Evolution}\label{sec:tcpnc-window}
\begin{figure}[tbp]
\begin{center}
\includegraphics[width=0.4\textwidth]{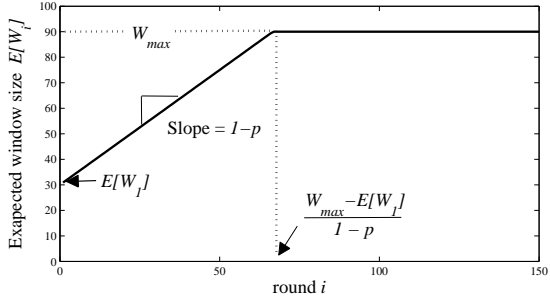}
\end{center}\vspace*{-.3cm}\caption{Expected window size for E2E where $W_{\max} = 90$, $E[W_1] = 30$, and $p = 0.1$. We usually assume $E[W_1] = 1$, however in this figure, $E[W_1] = 30$ to exemplify the effect of $E[W_1]$.}\label{fig:period}\vspace*{-.3cm}
\end{figure}
From Figure \ref{fig:td-tcpncrounds}, we observe that E2E-TCP/NC is able to maintain its window size despite experiencing losses. Furthermore, E2E is able to receive packets that would be considered out of order by TCP. Consequently, E2E can avoid closing its window due to erasures. This is because every packet that is linearly independent of previously received packets is considered to be ``innovative'' and is therefore acknowledged. As a result, E2E's window evolves differently from that of TCP, and can be characterized by a simple recursive relationship as
\begin{equation}\label{eq:tcpnc-w}
E[W_i] = E[W_{i-1}] + \frac{E[a_{i-1}]}{E[W_{i-1}]} = E[W_{i-1}] + (1-p).
\end{equation}
Note that the expected number of packets received in round $i$, $E[a_{i}] = (1-p)E[W_i]$. Once we take the maximum window size $W_{\max}$ into account, we have the following expression for E2E's expected window size:
\begin{equation}\label{eq:tcpnc-w2}
E[W_i] = \min(W_{\max}, E[W_1] + i(1-p)),
\end{equation}
where $i$ is the round number. $E[W_1]$ is the initial window size, and we set $E[W_1] = 1$. Figure \ref{fig:period} shows an example of the evolution of the E2E window using Equation (\ref{eq:tcpnc-w2}).

\subsubsection{Markov Chain Model}
\begin{figure*}[tbp]
\begin{center}
\includegraphics[width=0.68\textwidth]{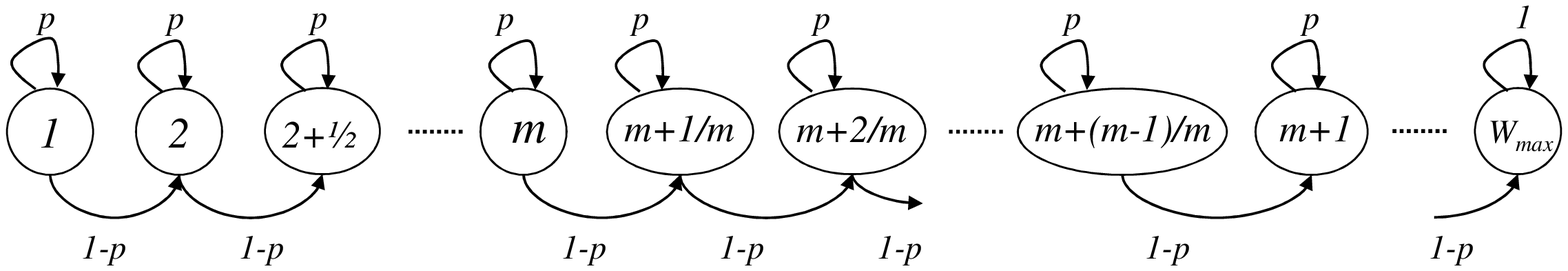}
\end{center}\vspace*{-.3cm}\caption{Markov chain for the E2E's window evolution.}\vspace*{-.4cm}\label{fig:markov}
\end{figure*}

The above analysis describes the expected behavior of E2E's window size. We can also describe the window size behavior using a Markov chain as shown in Figure \ref{fig:markov}. The states of this Markov chain represent the instantaneous window size (not specific to a round). A transition occurs whenever a packet is sent. We denote $S(W)$ to be the state representing the window size of $W$. Assume that we are at state $S(W)$. If a transmitted packet is received by the E2E receiver and acknowledged, the window is incremented by $\frac{1}{W}$; thus, we end up in state $S(W + \frac{1}{W})$. Note that this occurs with probability $(1-p)$. On the other hand, if the packet is lost, then the window stays at $S(W)$. This occurs with probability $p$. Thus, the Markov chain states represent the window size, and the transitions correspond to packet transmissions.

Note that $S(W_{\max})$ is an absorbing state of the Markov chain. As noted in Section \ref{sec:td-intuition}, E2E does not often experience a window shutdown, which implies that the network is in down-state. Thus, during the network up-state, E2E's window size is a non-decreasing, as shown in Figure \ref{fig:markov}. Therefore, given enough time, we will reach state $S(W_{\max})$ with probability equal to 1. Thus, we analyze the expected number of packet transmissions needed for absorption.

The \emph{transition matrix} $P$ and the \emph{fundamental matrix} $N = (I-Q)^{-1}$ of the Markov chain is given in Figure \ref{fig:transition}. The entry $N(S_1, S_2)$ represents the expected number of visits to state $S_2$ before absorption -- \ie we reach state $S(W_{\max})$ -- when we start from state $S_1$. Our objective is to find the expected number of packets transmitted to reach $S(W_{\max})$ starting from state $S(E[W_1])$ where $E[W_1]=1$. Thus, the partial sum of the first row entries of $N$ gives the expected number of packets transmitted until we reach the window size $W$. The expression for the first row of $N$ can be derived using cofactors:
\begin{equation}\label{eq:fundamental}
N(1,:) = \left[\frac{1}{1-p}, \frac{1}{1-p}, \cdots, \frac{1}{1-p}\right].
\end{equation}
Therefore, the expected number of packet transmissions $T(W)$ to reach a window size of $W \in [1, W_{\max}]$ is:
\begin{align}
T(W) &= \sum_{m = S(E[W_1]) = S(1)}^{S(W)} N(1, m) = \sum_{m=S(1)}^{S(W)} \frac{1}{1-p}\\
 &= \frac{S(W)-S(1)}{1-p} = \frac{W(W-1)}{2(1-p)}.\label{eq:R}
\end{align}
Equation (\ref{eq:R}) is due to the fact that there are $\sum_{m=1}^{W-1} m = \frac{W(W-1)}{2}$ states before state $S(W)$.

Note that $T(W)$ is equal to the area under the curve for rounds $i \in [0, \frac{W-E[W_1]}{1-p}]$ in Figure \ref{fig:period}. This corresponds to the sum of $E[W_i]$ (\ie the number of packets transmitted every round) as it increases from $E[W_1]=1$ to $W$.

\begin{figure}[tbp]
\[ P = \scriptsize
\left(
\begin{array}{cccccccccc}
\cellcolor[gray]{.8} p &\cellcolor[gray]{.8} 1-p &\cellcolor[gray]{.8}  0 & \cellcolor[gray]{.8} 0 &\cellcolor[gray]{.8}  0 &\cellcolor[gray]{.8} \cdots & \cellcolor[gray]{.8} 0 & 0  \\
\cellcolor[gray]{.8} 0 & \cellcolor[gray]{.8} p & \cellcolor[gray]{.8} 1-p &\cellcolor[gray]{.8}  0 &\cellcolor[gray]{.8}  0 &\cellcolor[gray]{.8}  \dotsb &\cellcolor[gray]{.8}  0 & 0  \\
\cellcolor[gray]{.8} 0 & \cellcolor[gray]{.8} 0 &\cellcolor[gray]{.8}  p & \cellcolor[gray]{.8} 1-p & \cellcolor[gray]{.8} 0&\cellcolor[gray]{.8}  \dotsb &\cellcolor[gray]{.8}  0 & 0  \\
\cellcolor[gray]{.8} \vdots & \cellcolor[gray]{.8}  &  \cellcolor[gray]{.8} & \cellcolor[gray]{.8} \ddots & \cellcolor[gray]{.8} \ddots&  \cellcolor[gray]{.8} \cdots &\cellcolor[gray]{.8} & \vdots \\
\cellcolor[gray]{.8} 0 &\cellcolor[gray]{.8}  0 & \cellcolor[gray]{.8} 0 & \cellcolor[gray]{.8} 0 & \cellcolor[gray]{.8} 0 &\cellcolor[gray]{.8} 0& \cellcolor[gray]{.8}  p & 1-p\\
0 & 0 & 0 & 0 & 0 & 0&0 & 1\\
\end{array}
\right)
\]\caption{The \emph{transition matrix} $P$ for the Markov chain in Figure \ref{fig:markov}. The shaded part of the matrix is denoted $Q$. Matrix $N = (I-Q)^{-1}$ is the \emph{fundamental matrix} of the Markov chain, and can be used to compute the expected rounds until the absorption state.}\vspace*{-.3cm}\label{fig:transition}
\end{figure}

\subsection{E2E-TCP/NC Throughput Analysis per Round}\label{sec:e2e-round}

Using the results in Section \ref{sec:tcpnc-window}, we derive an expression for the throughput. Once we have the expected value of the window size for any given round $i$, the throughput is straight forward. The throughput of round $i$, $\thru{i}$ is directly proportional to the window size $E[W_i]$, \ie
\begin{equation}\label{eq:nc-throughput-round}
\thru{i}  = \frac{1-p}{R\cdot SRTT}E[W_i] \text{ packets per second,}
\end{equation}
where $R$ is the redundancy factor of E2E-TCP/NC, and $SRTT$ is the ``effective'' round trip time. We shall formally define and discuss the effect of $R$ and $SRTT$ in the subsequent subsections.

We note that $\thru{i} \propto (1-p)E[W_i]$. At any given round $i$, we expect to see $E[W_i]$ packets transmitted by the E2E sender, and we expect $p E[W_i]$ packets to be lost. Thus, the E2E receiver can only acknowledge $(1-p) E[W_i]$ packets (or degrees of freedom), which results in the sender incrementing its window by $(1-p)$. Thus, $\thru{i} \propto (1-p)E[W_i]$ coincides with our intuition.

\subsubsection{Redundancy Factor $R$}\label{sec:redundancy}
The redundancy factor $R \geq 1$ is the ratio between the average rate at which linear combinations are sent to the receiver and the rate at which TCP's window progresses. For example, if the sender has 10 packets in its window, then TCP/NC transmits $10R$ linear combinations, unlike TCP which would send just 10 packets. This redundancy is necessary to (a) compensate for the losses within the network, and (b) match TCP's sending rate to the rate at which data is actually received at the receiver. References \cite{tcpnc}\cite{tcpnc2} introduce the redundancy factor with TCP/NC, and show that the optimal value is $R = \frac{1}{1-p}$.

It is important to discuss the effect of $R$ in the throughput. As noted in Equation (\ref{eq:nc-throughput-round}), the throughput is inversely proportional to the redundancy factor $R$. For example, for every 10 coded packets transmitted, there are only $\frac{10}{R}$ original data packets from the source. Thus, given any number of coded packets transmitted by the sender, the receiver only needs $\frac{1}{R}$-fraction of them and the rest $1 - \frac{1}{R}$-fraction of the packets are redundant. Therefore, the throughput at any given round $i$ is inversely proportional to the redundancy factor. Specifically, $\thru{\text{i}} \propto \frac{E[W_i]}{R}$.

The redundancy factor $R$ should be chosen with some care. By Equation (\ref{eq:nc-throughput-round}), it may seem that setting $R = 1$ would maximize throughput. However, setting $R < \frac{1}{1-p}$ causes significant performance degradation, since network coding can no longer fully compensate for the losses which may lead to window closing for E2E. To maximize throughput, an optimal value of $R = \frac{1}{1-p}$ should be chosen. However, $R$ need not be set to the optimal value for E2E-TCP/NC to work. Setting $R > \frac{1}{1-p}$ means that network coding may over-compensate for the losses within the network; thus, introducing more redundant packets than necessary. Thus, we assume that $R \geq \frac{1}{1-p}$.

\subsubsection{Effective Round Trip Time $SRTT$}\label{sec:srtt}

$SRTT$ is the round trip time estimate that TCP maintains by sampling the behavior of packets sent over the connection. It is denoted $SRTT$ because it is often referred to as ``smoothed'' round trip time as it is obtained by averaging the time for a packet to be acknowledged after the packet has been sent. We note that, in Equation (\ref{eq:nc-throughput-round}), we use $SRTT$ instead of $RTT$ because $SRTT$ is the ``effective'' round trip time E2E experiences.

For E2E operating in lossy networks, $SRTT$ is often greater than $RTT$. This can be seen in Figure \ref{fig:example}. The first coded packet ($\mathbf{p_1 + p_2 + p_3}$) is received and acknowledged ($\mathbf{seen(p_1)}$). Thus, the sender is able to estimate the round trip time correctly; resulting in $SRTT \approx RTT$. However, the second coded packet ($\mathbf{p_1 + 2p_2 + p_3}$) is lost. As a result, the third packet ($\mathbf{p_1 + 2p_2 + 2 p_3}$) is used to acknowledge the second degree of freedom ($\mathbf{seen(p_2)}$). We note that in our model, we assume for simplicity that the time needed to transmit a packet is much smaller than RTT (Section \ref{sec:model}); thus, despite the losses, our model would result in $SRTT \approx RTT$. However, in practice, depending on the size of the packets, the transmission time may not be negligible. This results in a longer round trip time estimate, which can be characterized as described below.

We define $t_p$ to be the time to transmit a packet. Then, the sender expects to receive an ACK of a packet after $SRTT$ time units, where
\begin{align}
SRTT & = \sum_{i = 0}^{\infty} (RTT + i\cdot t_p)p^i (1-p)\\
& = RTT + t_p \frac{p}{1-p}\label{eq:srtt}.
\end{align}
For simplicity, Equation (\ref{eq:srtt}) does not take into account the ``edge effect'' of packets that are waiting to be acknowledged across rounds. As the window size grows, the edge effect can safely be ignored.

\subsection{E2E-TCP/NC Average Throughput}\label{sec:e2e-average}

Taking Equation (\ref{eq:nc-throughput-round}), we can average the throughput over $n$ rounds to obtain the average throughput for E2E-TCP/NC.
\begin{align}
\thru{e2e} &= \frac{1}{n}\sum_{i=1}^n \frac{1-p}{R \cdot SRTT} E[W_i]\\
&=\frac{1-p}{nR \cdot SRTT} \sum_{i=1}^n \min(W_{\max}, E[W_1] + i (1-p))\\
&= \frac{1-p}{nR \cdot SRTT} \cdot f(n),\label{eq:nc-final-longterm}
\end{align}
where
\begin{align*}
f(n) &=
\begin{cases} nE[W_1] + (1-p) \frac{n(n+1)}{2} \text{\hspace*{.5cm} for $n \leq r^*$}\\
nW_{\max} - r^*(W_{\max} - E[W_1]) + (1-p) \frac{r^*(r^*-1)}{2}\\
\text{\hspace*{1cm} otherwise}
\end{cases}\\
r^* &= \frac{W_{\max} - E[W_1]}{1-p}.
\end{align*}
Note that as $n \rightarrow \infty$, the average throughput $\thru{e2e} \rightarrow \frac{1-p}{R \cdot SRTT} W_{\max}$.

\section{Network in down-state}\label{sec:down}

In this section, we consider the effect of $p_d$, the probability of network being in down-state at any given round, during which the sender is unable to transfer data to the receiver (for both TCP and E2E). Note that the average throughput analysis in Sections \ref{sec:td} and \ref{sec:tcpnc} gives the throughput during when the network is in up-state. Thus, the average throughput would depend on 1) the average number of rounds during the network up-state $E[r_{up}]$, and 2) the average length of down-state periods $E[r_{down}]$.

\begin{figure}[tb]
\begin{center}
\subfloat[TCP]{\label{fig:to-tcp}\includegraphics[width=.45\textwidth]{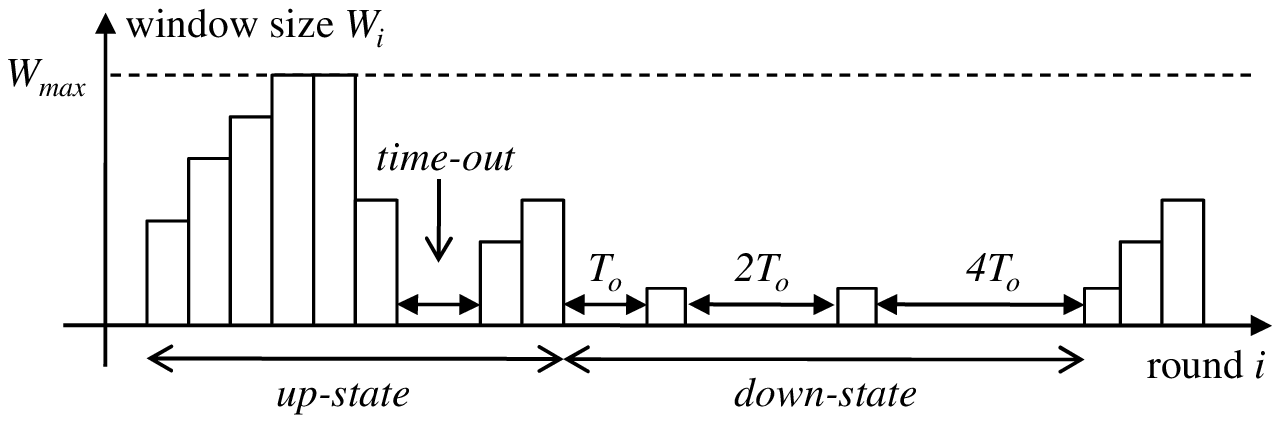}}\\
\vspace*{-.2cm}
\subfloat[TCP/NC]{\label{fig:to-tcpnc}\includegraphics[width=.45\textwidth]{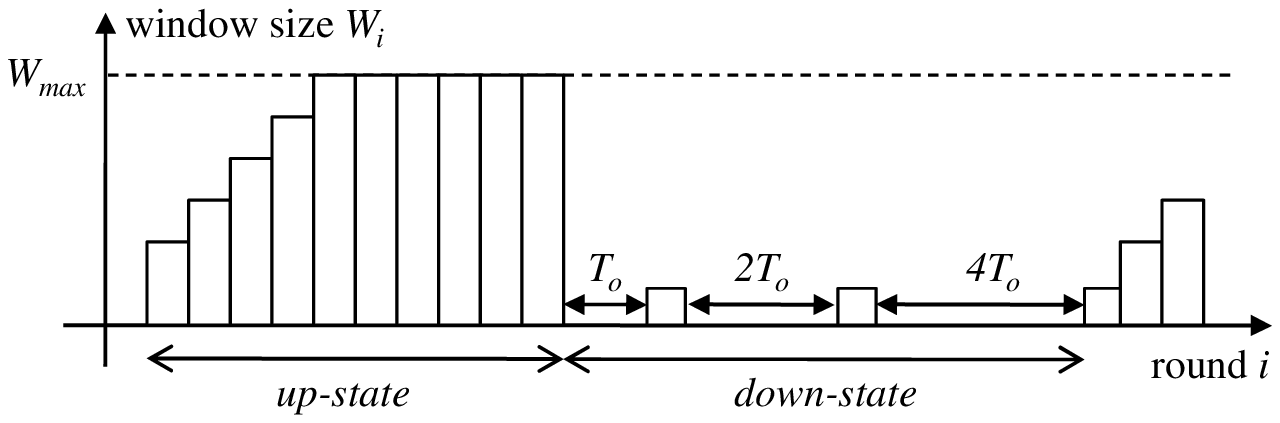}}
\end{center}\vspace*{-.3cm}
\caption{The effect of time-out events and maximum window $W_{\max}$: When a time-out occurs, both TCP and E2E experience zero throughput. The maximum window size $W_{\max}$ limits both TCP and E2E. The performance difference comes from the frequency and the duration of TCP or E2E achieving the window size of $W_{\max}$.}\label{fig:td}\vspace*{-.3cm}
\end{figure}\label{fig:to}

The average number of rounds during a network upstate is $E[r_{up}] = \sum_{i=0}^{\infty} i (1-p_d)^i p_d = \frac{1-p_d}{p_d}$. The average length of down-state period is $E[r_{down}] = \sum_{i=0}^{\infty} i p_d^i (1-p_d) = \frac{p_d}{1-p_d}$.

%

Taking into account both TD and TO events, the average throughput of TCP and E2E can be summarized as below. The long term average throughput of TCP is given in Equation (\ref{eq:tcp-final1}). The long term average throughput of E2E is given in Equation (\ref{eq:nc-final1}). As discussed in Section \ref{sec:e2e-average}, $\thru{e2e}$ depends on the number of rounds $n$; thus, the length of $E[r_{up}]$ affects its performance.

\begin{figure*}[tbp]
\begin{align}
\thru{tcp} &=
\begin{cases}
\min\left(\frac{W_{\max}}{RTT}, \frac{1-p}{p}\frac{1}{RTT\left(\frac{5}{3} + \sqrt{-\frac{1}{18} + \frac{2}{3}\frac{1-p}{p}} + \mathbf{P}(\text{TO}|E[W])E[\text{duration of TO period}]\right)} \right) &\text{ when network is in up-state;}\\
0 &\text{ when network is in down-state.}
\end{cases}\label{eq:tcp-final1}\\
\thru{e2e}  &=
\begin{cases}
 \frac{1-p}{nR \cdot SRTT} \cdot f(n) & \text{ when network is in up-state;}\\
 0 &\text{ when network is in down-state,}
\end{cases}\label{eq:nc-final1}\\
&\text{ where } f(n) = \begin{cases} nE[W_1] + (1-p) \frac{n(n+1)}{2} &\text{for $n \leq \frac{W_{\max} - E[W_1]}{1-p}$;}\\
nW_{\max} - \frac{(W_{\max} - E[W_1])^2}{1-p} + \frac{1}{2}(W_{\max} - E[W_1])(\frac{W_{\max} - E[W_1]}{1-p}-1) &\text{otherwise}.
\end{cases}
\end{align}\vspace*{-.3cm}
\end{figure*}



\section{Simulation results}\label{sec:simulations}

\begin{figure}[tbp]
\begin{center}
\includegraphics[width=0.30\textwidth]{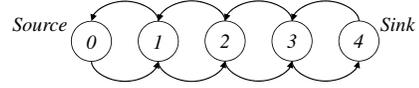}
\end{center}\vspace*{-.3cm}\caption{Network topology for the simulations.}\label{fig:setup}\vspace*{-.5cm}
\end{figure}

In this section, we use simulations to verify that our analysis captures the behavior of both TCP and E2E. We use NS-2 (Network Simulator \cite{ns}) to simulate TCP and E2E-TCP/NC where TCP-Vegas is used as the underlying TCP protocol. We use the implementation of E2E from \cite{tcpnc2}. The network topology for the simulation is shown in Figure \ref{fig:setup}. All links, in both forward and backward paths, are assumed to have a bandwidth of 1 Mbps, a propagation delay of 100 ms, a buffer size of 200, and a erasure rate of $q$. Each packet transmitted is assumed to be 8000 bits (1000 bytes). We set $W_{\max}= 90$ packets for all simulations. We fix the redundancy factor $R = 1.25$ for all $q$. We note that the optimal redundancy factor varies with $q$. In addition, time-out period $T_o = \frac{3}{RTT} = 3.75$ rounds long (3 seconds).

\begin{figure*}
\begin{center}
\subfloat[Throughput]{\includegraphics[width=0.47\textwidth]{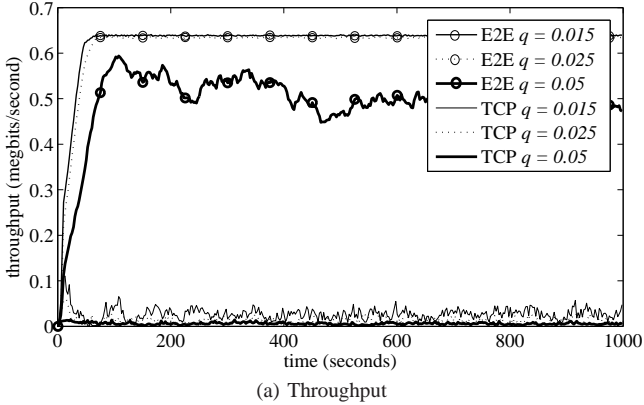}\label{fig:bw}}\hspace*{.5cm}
\subfloat[Window size]{\includegraphics[width=0.47\textwidth]{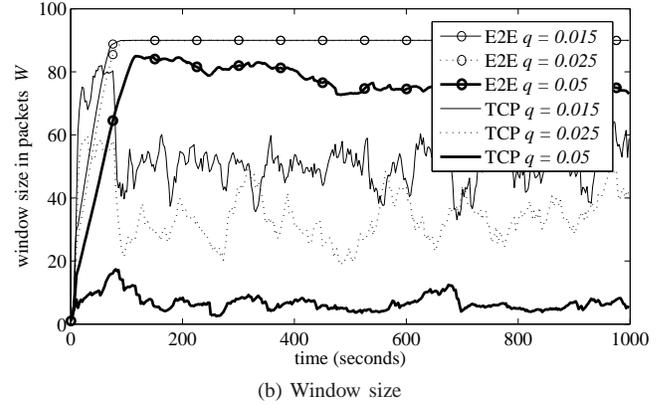}\label{fig:cw}}
\end{center}\vspace*{-.3cm}\caption{Throughput and congestion window size of TCP and E2E with varying link erasure probability $q$. Each curve is an average of 30 independent NS-2 simulation results.}\label{fig:bcw}\vspace*{-.7cm}
\end{figure*}

To compare the performance of TCP and E2E, we average the performance over 30 independent runs of the simulation, each of which is 1000 seconds long. We vary the per link probability of erasure, $q$. We consider $q =0.015$, 0.025, and $0.05$. Note that since there are in total four links in the path from node 0 to node 4, the probability of packet erasure is $p = 1-(1-q)^4$. Therefore, the corresponding $p$ values are 0.0587, 0.0963, and 0.1855.

First, we show the throughput benefits of E2E over TCP in Figure \ref{fig:bcw}. As our analysis predicts, E2E sustains its high throughput despite the erasures present in the network. We observe that TCP may close its window due to triple-duplicates ACKs; however, E2E is more resilient to such erasure patterns. Therefore, E2E is able to increment its window consistently even with erasures and achieve a window size of $W_{\max}$ sooner than that of TCP. In Figure \ref{fig:cw}, we observe that E2E converges to $W_{\max}=90$ very rapidly for $q =0.015$, and 0.025. More importantly, E2E is able to \emph{maintain} the window size of 90 even under lossy conditions when standard TCP is unable to (resulting in the window fluctuation in Figure \ref{fig:cw}).

We note that for $q = 0.05$, the throughput behavior is not as steady. This is due to the increase in loss rate, \ie $p = 0.1855$. Given that $p = 0.1855$ and link capacities 1 Mbps, the effective maximum throughput we can achieve is $(1-0.1855) = 0.8145$ Mbps. Since each packet is 8000 bits and RTT = 0.8 seconds, we can expect the throughput to be at most $\frac{0.8145 \times 10^6}{8000/0.8} = 81.45$ packets per second, which is less than $W_{\max} = 90$. As a result, the bottleneck is not $W_{\max}$ but the link capacity for $q = 0.05$. This is consistent with the performance we see in Figure \ref{fig:cw}.

\begin{figure}[tbp]\vspace*{.3cm}
\begin{center}
\includegraphics[width=0.46\textwidth]{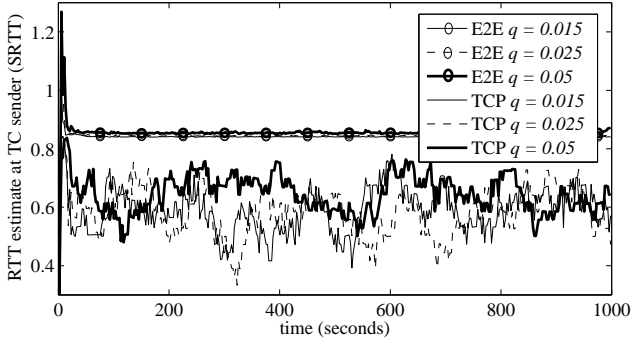}\label{fig:rtt}
\end{center}\vspace*{-.3cm}\caption{The round-trip time estimate (SRTT) at the TCP sender (Node 0).}\label{fig:rtt}\vspace*{-.6cm}
\end{figure}

An interesting observation is that, TCP achieves a moderate average window size (depending on $q$, 20-60 packets) while E2E achieves average window size of 90 (where $W_{\max} = 90$). However, the average throughput of E2E is much higher than that of TCP's, as shown in Figure \ref{fig:bw} and Table \ref{tb:thru}. This is because, before closing its window (TD or TO event), the TCP sender waits for a certain period of time, called \emph{retransmission timeout period}. This retransmission timeout period is approximately set to $2\cdot RTT$. During this retransmission timeout period, the TCP sender maintains its window size, despite the fact that it is idle and waiting for acknowledgements. Thus, for TCP, the average window size may be much larger than the average throughput. On the other hand, E2E does not experience TD or TO events as often; thus, the window size commensurate the average throughput. Although we did not explicitly take retransmission timeout into consideration in the TCP analysis in Section \ref{sec:td}, the retransmission timeout is not difficult to incorporate. For every TD or TO event, TCP idles for two extra round-trip time, which can be incorporated by letting $E[\Delta] \leftarrow E[\Delta] + 2$ and $E[\text{duration of TO period}] \leftarrow E[\text{duration of TO period}] + 2$.

\begin{figure*}[tbp]
\begin{center}
\subfloat[E2E-TCP/NC]{\includegraphics[width=0.45\textwidth]{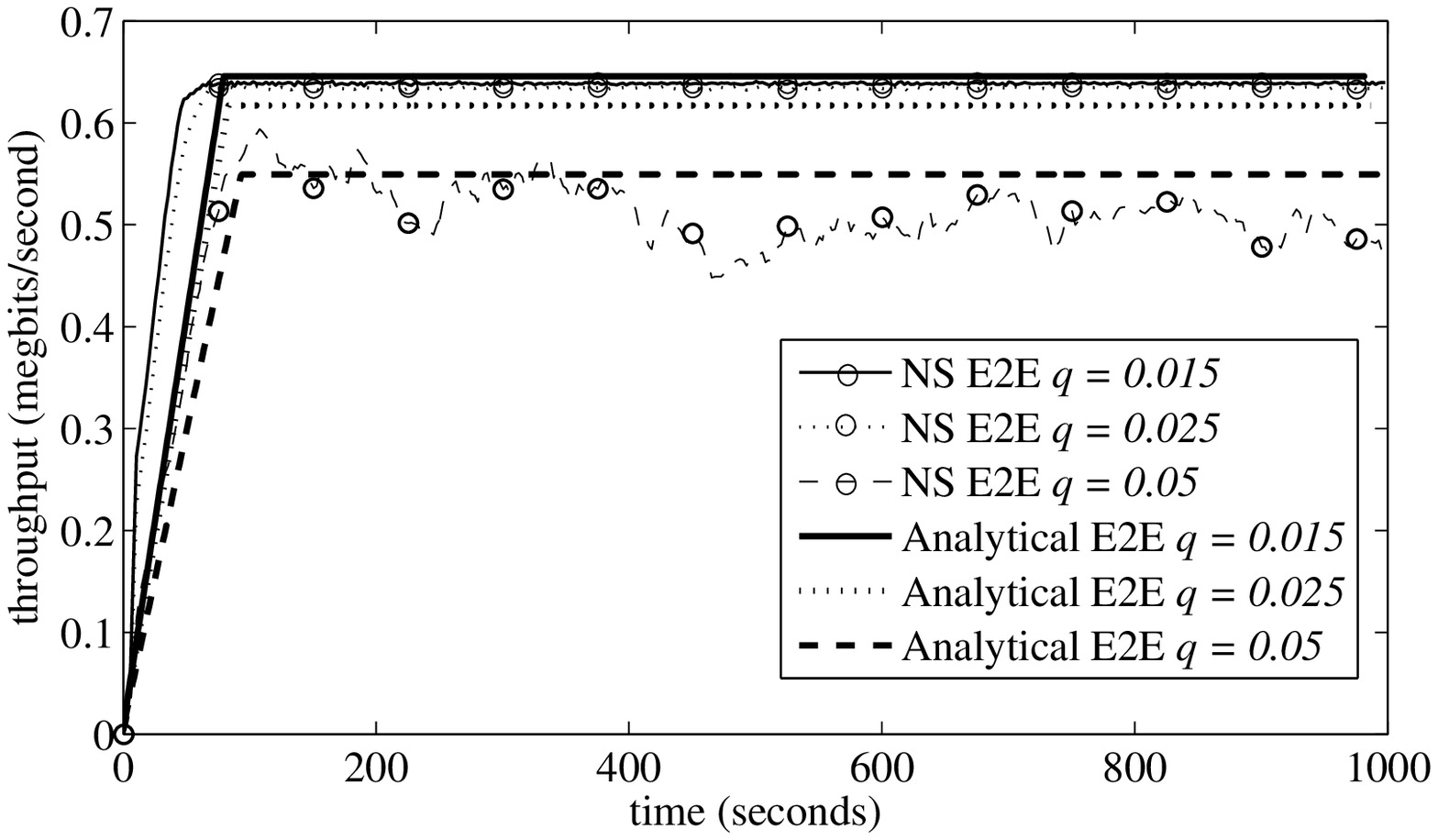}}\hspace*{.5cm}
\subfloat[TCP]{\includegraphics[width=0.47\textwidth]{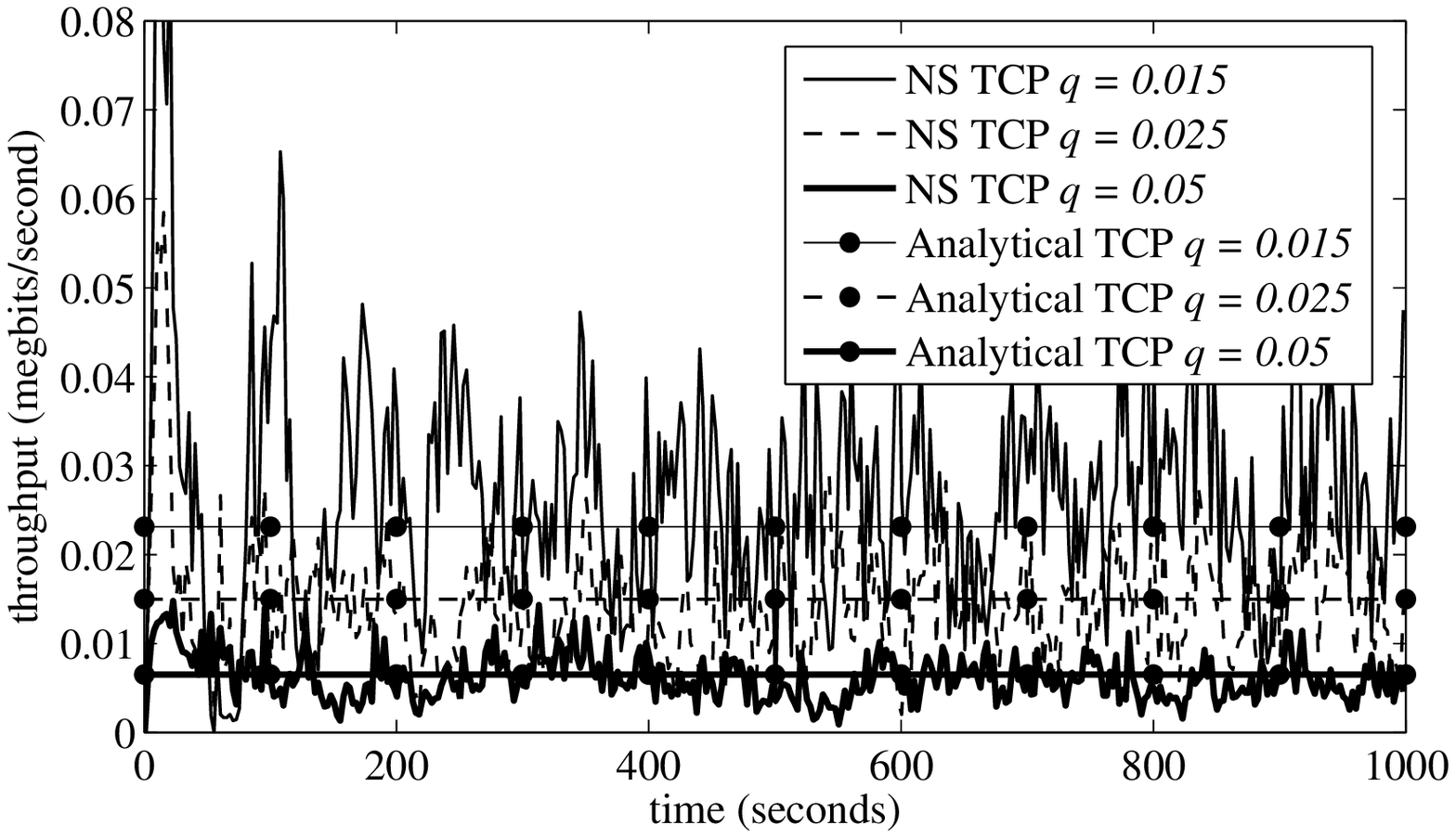}}
\end{center}\vspace*{-.2cm}\caption{Average throughput of E2E and TCP with varying link erasure probability $q$. `NS E2E' or `NS TCP' represent the average throughput achieved over 30 independent NS-2 simulations; `Analytical E2E' and `Analytical TCP' are computed using Equations (\ref{eq:nc-throughput-round}) and (\ref{eq:tcp}), respectively. For `Analytical E2E', we assume that $R = 1.25$, $E[W_1] = 1$, and $SRTT$ is obtained from the simulation results, as shown in Table \ref{tb:thru}.}\label{fig:bw_tcp_nc}
\end{figure*}

As described in Sections \ref{sec:td-intuition} and \ref{sec:srtt}, E2E masks errors by translating losses as longer RTT. For E2E, if a specific packet is lost, the next subsequent packet received can ``replace'' the lost packet; thus, allowing the receiver to send an ACK. Therefore, the longer RTT estimate takes into account the delay associated with waiting for the next subsequent packet at the receiver. In Figure \ref{fig:rtt} and Table \ref{tb:thru}, we verify that this is indeed true. In the simulation, we have a round trip time of $8\cdot 100 = 800$ ms. TCP, depending on the ACKs received, modifies its RTT estimation; thus, due to random erasures, TCP's RTT estimate fluctuates significantly. On the other hand, E2E is able to maintain a consistent estimate of the RTT; however, is slightly above the actual 800 ms.

\begin{table*}[tbp]
\caption{The average simulated or predicted long-term throughput of TCP and E2E in megabits per second (Mbps). ` simulation' and `TCP simulation' are averaged over 30 trials for 1000 seconds each. `Analytical E2E' is calculated using Equation (\ref{eq:nc-final-longterm}) with $\lfloor n\cdot SRTT \rfloor = 1000$. `TCP analysis' is computed using Equation (\ref{eq:tcp}) while taking into account the retransmission timeout period. }\label{tb:thru}
\centering
\begin{tabular}{|c|c|c|c|c|c|c|}
\hline
$q$ & $p = 1-(1-q)^4$ & Average E2E SRTT & E2E simulation & E2E analysis ($\lfloor n \cdot SRTT \rfloor = 1000$) & TCP simulation & TCP analysis \\
\hline
0.015 & 0.0587 & 0.8396 & 0.6242 & 0.6202 & 0.0264 & 0.0231\\
0.025 & 0.0963 & 0.8434 & 0.6161 & 0.5917 & 0.0136 & 0.0150\\
0.05 & 0.1855 & 0.8540 & 0.5200 & 0.5243 & 0.0067 & 0.0065\\
\hline
\end{tabular}\vspace*{-.2cm}
\end{table*}

Finally, we examine the accuracy of our analytical model in predicting the behavior of TCP and E2E. First, note that our analytical model of window evolution (shown in Equation (\ref{eq:tcpnc-w2}) and Figure \ref{fig:period}) demonstrates the same trend as that of the window evolution of E2E NS-2 simulations (shown in Figure \ref{fig:cw}). Second, we compare the actual NS-2 simulation performance to the analytical model. This is shown in Figure \ref{fig:bw_tcp_nc} and Table \ref{tb:thru}. From the results, we observe that Equations (\ref{eq:nc-throughput-round}) and (\ref{eq:tcpnc-w2}) predict well the trend of E2E's throughput and window evolution, and provides a good estimate of E2E's performance. Furthermore, our analysis predicts the average TCP behavior very well. In Figure \ref{fig:bw_tcp_nc} and Table \ref{tb:thru}, we see that Equation (\ref{eq:tcp}) is consistent with the NS-2 simulation results even for large values of $p$. Therefore, both simulations as well as analysis support that E2E is resilient to erasures; thus, better suited for reliable transmission over unreliable networks, such as wireless networks.


\section{Conclusions}\label{sec:conclusions}

We have presented an analytical study and compared the performance of TCP and E2E-TCP/NC. Our analysis characterizes the throughput of TCP and E2E as a function of erasure rate, round-trip time, maximum window size, and duration of the connection. We showed that network coding, which is robust against erasures and failures, can prevent TCP's performance degradation often observed in lossy networks. Our analytical model shows that TCP with network coding has significant throughput gains over TCP. E2E is not only able to increase its window size faster but also to maintain a large window size despite losses within the network; on the other hand, TCP experiences window closing as losses are mistaken to be congestion. Furthermore, NS-2 simulations verify our analysis on TCP's and E2E's performance. Our analysis and simulation results both support that E2E is robust against erasures and failures. Thus, E2E is well suited for reliable communication in lossy wireless networks.
\bibliography{References}
\bibliographystyle{IEEEtran}

\end{document}